\documentclass[prl,aps,showpacs,twocolumn,preprintnumbers,amsmath,amssymb,superscriptaddress,longbibliography]{revtex4-1}\usepackage{standalone}
\usepackage{nccfoots}
\usepackage{comment}
\usepackage{graphicx}
\usepackage{braket}
\usepackage{bbold}
\usepackage{color}
\usepackage{svg}
\usepackage{physics}
\usepackage{float}
\usepackage{comment}
\usepackage[normalem]{ulem}

\definecolor{mygreen}{rgb}{0,0.5,0}
\definecolor{myblue}{rgb}{0,0,0.75}
\definecolor{mymagenta}{cmyk}{0,1,0,0.12}
\usepackage[colorlinks=true,
   linkcolor=blue,
   citecolor=blue,
   urlcolor =blue]{hyperref}

\newcommand{\citeSM}{\cite[{\tiny SM}\kern-0.3em][]{SM}}

\usepackage{amsmath,amsfonts,amssymb,amsthm, bbm,braket}
\usepackage{subfigure}

\expandafter\let\csname equation*\endcsname\relax

\expandafter\let\csname endequation*\endcsname\relax

\usepackage{amsmath}
\begin{document}

\title{Spectroscopy of edge and bulk collective modes in fractional Chern insulators}

\author{F. Binanti}
 \affiliation{Univ. Grenoble-Alpes, CNRS, LPMMC, 38000 Grenoble, France}

 \author{N. Goldman}
 \affiliation{CENOLI, Université Libre de Bruxelles, CP 231, Campus Plaine, B-1050 Brussels, Belgium}

 \author{C. Repellin}
 \affiliation{Univ. Grenoble-Alpes, CNRS, LPMMC, 38000 Grenoble, France}

\date{\today}

\begin{abstract}
The exploration of atomic fractional quantum Hall (FQH) states is now within reach in optical-lattice experiments. While ground-state signatures have been observed in a system realizing the Hofstadter-Bose-Hubbard model in a box [Leonard et al., Nature 2023], how to access hallmark low-energy collective modes remains a central open question in this context.
We introduce a spectroscopic scheme based on two interfering Laguerre-Gaussian beams, which transfer a controlled angular momentum and energy  to the system. The edge and bulk responses to the probe are detected through local density measurements, by tracking the transfer of atoms  between the bulk and the edge of the FQH droplet. This detection scheme is shown to simultaneously reveal two specific signatures of FQH states:~their chiral edge branch and their bulk magneto-roton mode. We numerically benchmark our method by considering few bosons in the $\nu\!=\!1/2$ Laughlin ground state of the Hofstadter-Bose-Hubbard model, and demonstrate that these signatures are already detectable in realistic systems of two bosons, provided that the box potential is larger than the droplet. Our work paves the way for the detection of fractional statistics in cold atoms through edge signatures.
\end{abstract}

\maketitle

{\it Introduction --- }
The interplay of topology and interactions leads to fascinating phases of matter, such as the fractional quantum Hall (FQH) states, which host fractionalized anyonic excitations. The progress in engineering artificial gauge fields~\cite{Dalibard_review, Goldman_review} and topological bands~\cite{cooper2019topological} has raised the hope of realizing FQH states of ultracold atoms. Specifically, realistic protocols for the preparation of FQH states were proposed~\cite{popp2004adiabatic, cooper2013reaching, yao2013realizing, grusdt2014topological, he-PhysRevB.96.201103, repellin2017pairing, motruk2017preparation, hudomal2019bosonic, michen2023adiabatic}, based on the quasi-adiabatic evolution of a small ensemble of neutral atoms loaded into an optical lattice. In this context, extracting the universal signatures of FQH states is an important goal and great challenge; theoretical proposals in this direction have focused on the measurement of the Hall response~\cite{repellin2019detecting, repellin2020drift, motruk2020drift, cian2021MBChern}, central charge~\cite{palm2022snapshot}, and the anyonic properties of quasiholes~\cite{cooper2015statistics, raciunas2018fractional, umucalilar2018time, macaluso2020charge, botao2022Scipost, li2023dynamics}. Recently, a two-particle bosonic Laughlin state was identified in an optical lattice~\cite{leonard}, where local density measurements permitted the observation of key bulk signatures, including a nearly quantized Hall conductivity and vortex-like correlations.

Edge states are a fundamental hallmark of topological matter. In FQH systems, they form one-dimensional conduction channels~\cite{Wen1990ChiralLL}, which are responsible for a wealth of quantum coherent phenomena in mesoscopic systems~\cite{Bauerle_2018}.
Despite the success of chiral Luttinger liquid theory in capturing these phenomena, microscopic details such as boundary effects can deeply affect the low-energy picture~\cite{macaluso2017hardwall, fern2017hardconfinement, dong2018edge, carusotto2020NonLinear, oblak2023anisotropic, kjall2012edge, luo2013edge}.

Another key feature of FQH phases is their bulk collective excitations called magneto-roton mode (MRM), which can be viewed as a density modulation of the ground state~\cite{GMP1985, GMP1986}. Owing to the incompressibility of FQH phases, the MRM is gapped; its softening marks the transition to a Wigner crystal. While it is absent in integer quantum Hall phases, the MRM generically appears in all FQH phases.

Thanks to the local probes accessible in optical-lattice experiments, the realization of the FQH effect in quantum gases may provide an opportunity to reveal the rich phenomenology of FQH edge and bulk modes. Promisingly, resolving individual edge states would provide a marker of topological order, permitting the unambiguous detection of non-Abelian FQH states~\cite{Moore1991Nonabelion, wen1993nonabelian, milovanovic1996paired}. Moreover, the maximal angular momentum of the MRM could serve as an additional topological marker to distinguish FQH phases~\cite{KamillaPRB1996, rodriguezPRB2012, yangPRL2012, yangPRB2013, jolicoeur2017}. While edge properties have been detected in weakly interacting cold atom settings~\cite{Mancini2015edge, Stuhl_2015, Chalopin_2020, braun2023realspace, yao2023observation}, how to extract and resolve the low-energy spectrum of FQH states remains an open question.

\begin{figure}[t]
\begin{minipage}[b]{\linewidth}
\centering
\includegraphics[width=1.0\linewidth]{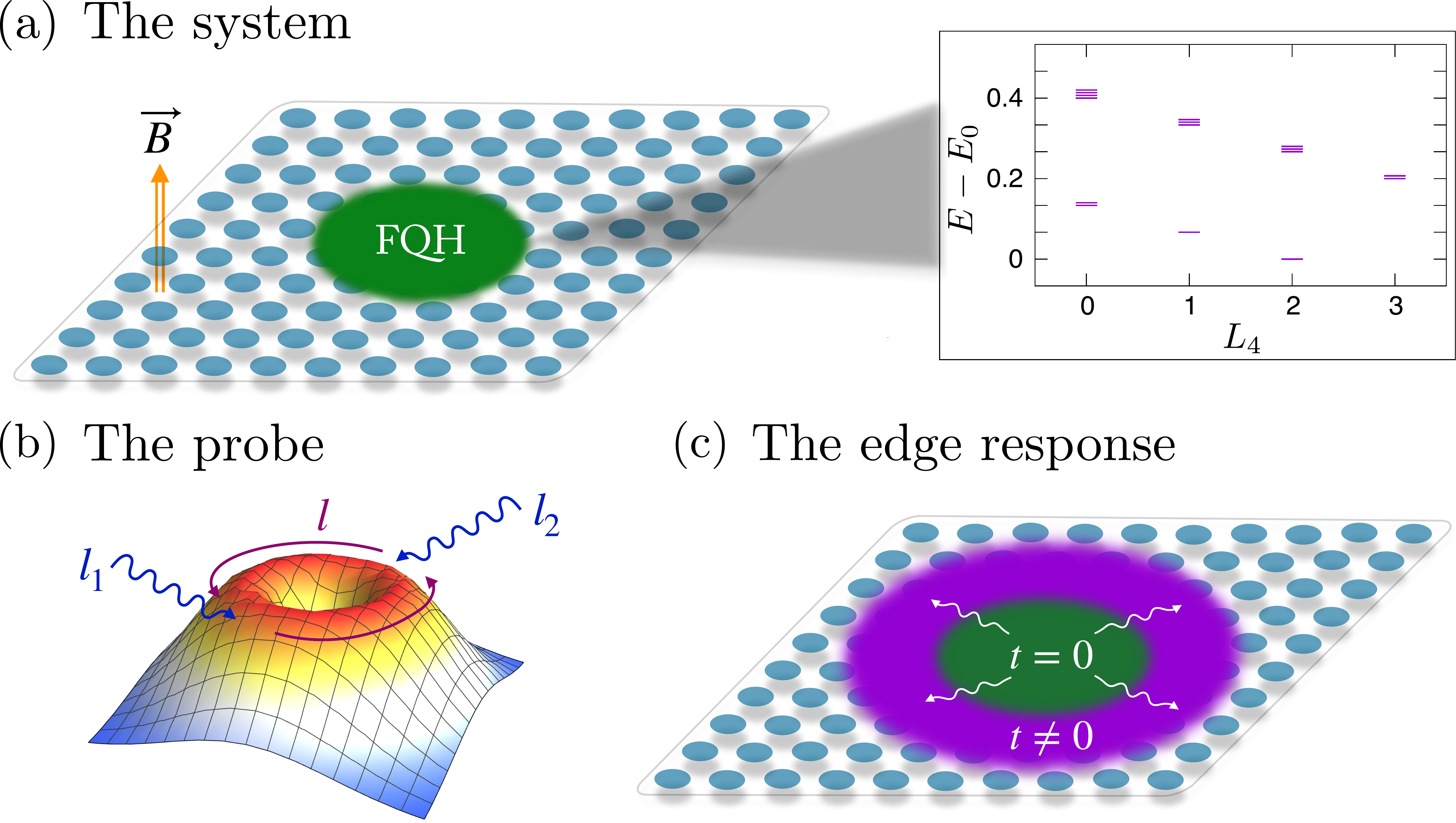}
\end{minipage}
\caption{(a) Fractional quantum Hall (FQH) droplet on the Hofstadter-Bose-Hubbard lattice, with a sketch of the expected edge spectrum. (b) Spatial shape of the Laguerre-Gaussian (LG) laser field acting on the atoms, realized by interfering two beams with angular momentum $l_1$ and $l_2$, and resulting in a transfer of angular momentum $l = l_1 - l_2$. (c) Response of the FQH droplet to the LG beams at resonance: edge states are populated, resulting in a detectable density increase at the droplet's edge.}
\label{FigureOne}
\end{figure}

In this work, we develop a spectroscopy protocol to extract the edge and bulk collective modes of FQH states in ultracold gases. Our proposal builds on Ref.~\cite{goldman2012edge} and is summarized in Fig.~1: two interfering Laguerre-Gaussian (LG) lasers transfer an angular momentum $l$ and energy $\hbar \omega$ to the system prepared in the FQH ground state of a lattice Hamiltonian. The absorption resonance is subsequently measured in-situ through local density measurements, by monitoring the transfer of atoms between the bulk and edge. We numerically benchmark our protocol using the experimentally realized~\cite{Tai2017, leonard} Hofstadter-Bose-Hubbard model, which supports a Laughlin $1/2$ ground state~\cite{sorensen-PhysRevLett.94.086803, hafezi-PhysRevA.76.023613, gerster-PhysRevB.96.195123}. By calculating the coupling matrix elements of our probe, we obtain the angular momentum resolved absorption spectrum and show that the characteristic edge and bulk collective modes are present in systems with as few as two particles. Interestingly, the LG matrix elements have an approximate selection rule corresponding to the emergent continuous rotation symmetry of low-energy states. This property allows us to extract the edge and bulk modes even when they are energetically mixed. Upon approaching the experimental configuration of Ref.~\cite{leonard}, the walls of the confining box nearly coincide with the edge of the FQH droplet, leading to a gapped edge mode lying higher in energy than the first bulk excitation. The explicit time-dependent simulation of our protocol shows that the time necessary to detect a measurable transfer of density from bulk to edge is compatible with current experimental constraints. Our protocol could apply to other models of FQH states of ultracold atoms, including rotating atomic traps~\cite{yao2023observation}.

{\it Model --- } We consider the Hofstadter-Bose-Hubbard (HBH) model, describing bosons hopping on the Harper-Hofstadter lattice~\cite{hofstadter1976energy} and interacting through an on-site Hubbard interaction of strength $U \!>\! 0$,
\begin{align}
    \hat H &= -J\sum_{m,n} \bigg ( \hat b^\dagger_{m+1,n}\hat b_{m,n}e^{i2\pi\alpha n} + \hat b^\dagger_{m,n+1}\hat b_{m,n} + \text{h.c.} \bigg ) \notag \\
    & + \frac{U}{2}\sum_{m,n} \hat b^\dagger_{m,n} \hat b_{m,n} \left( \hat b^\dagger_{m,n} \hat b_{m,n} - 1 \right)\notag \\
    & + \sum_{m,n} V(m,n) \hat b^\dagger_{m,n} \hat b_{m,n} ,
\label{HofstadterHamiltonian}
\end{align}
where the operator $\hat b_{m,n}$ ($\hat b^\dagger_{m,n}$) destroys (creates) a boson at site $(m,n)$, $J$ is the tunneling energy, $\alpha$ is the magnetic flux density, and $V(m,n) = V_0 \left((m-m_0)^2 + (n-n_0)^2\right)$ is a confining potential ($m_0,n_0$ are the coordinates of the lattice center). Throughout the paper, we work in the strong interaction regime $U\gg J\!=\!1$, and use hardcore bosons in the numerics unless otherwise stated.
For $\alpha < 1/3$, this model hosts a fractional Chern insulator ground state~\cite{sorensen-PhysRevLett.94.086803, hafezi-PhysRevA.76.023613, gerster-PhysRevB.96.195123}, which is a lattice analog of the $\nu=1/2$ bosonic Laughlin state. A cold-atom implementation of this model was realized using two bosons in a box potential~\cite{Tai2017, leonard}, revealing signatures of the Laughlin FQH state~\cite{leonard}.

\begin{figure}[t]
\centering
\includegraphics[width=\linewidth]{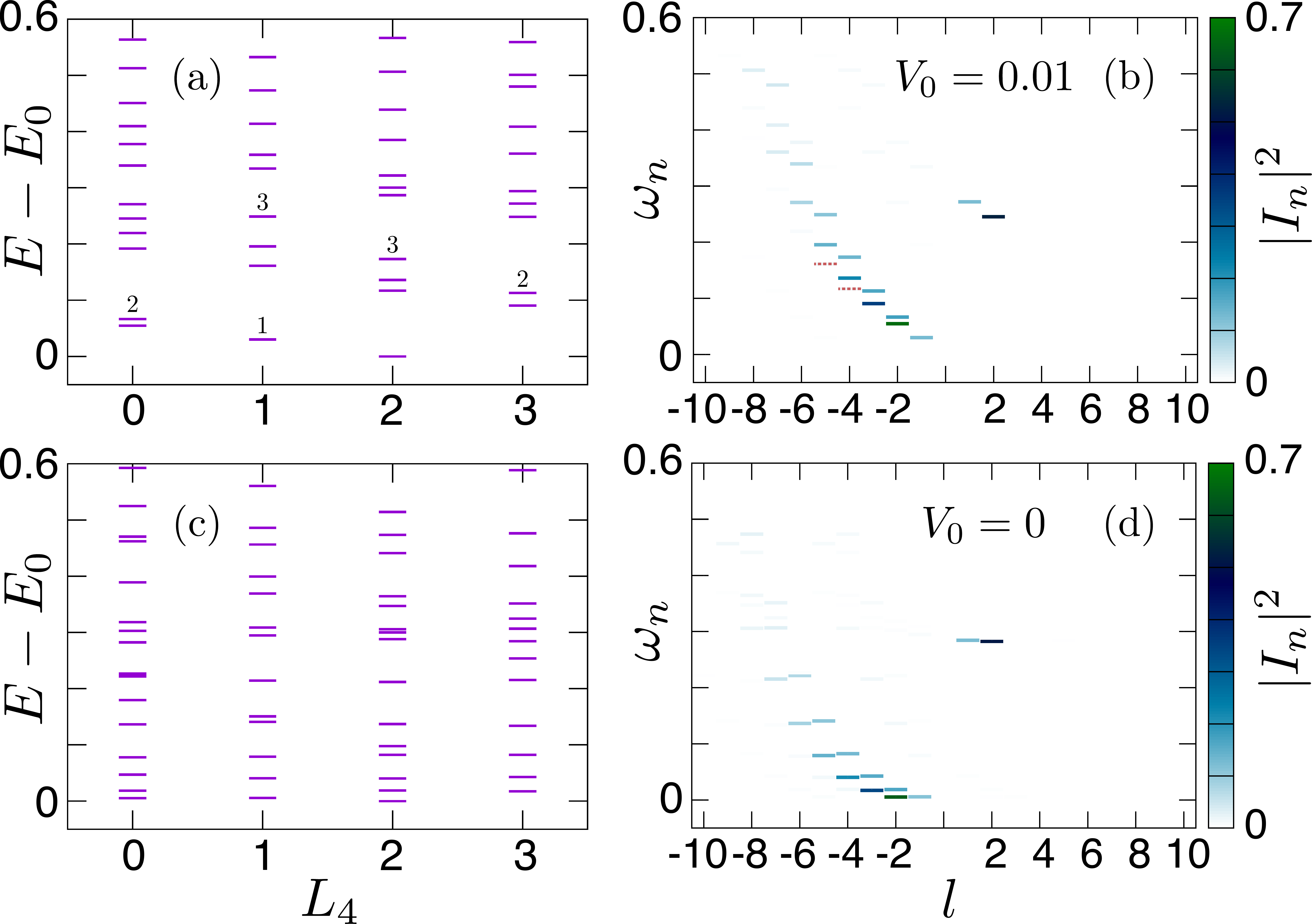}
\caption{Low-energy spectrum (a, c) and corresponding coupling matrix elements $I_n$ (b, d) for a system of $2$ hardcore bosons in the HBH model at flux density $\alpha = 1/8$, in a 10x10 box, with (a, b) or without (c, d) a harmonic potential of strength $V_0 = 0.01$.  $\omega_n$ is the energy difference between excited and ground states. Unlike the energy spectrum, the matrix elements always distinguish edge and bulk modes, regardless of the presence of a harmonic potential. The numbers in (a) indicate the number of states in each ``cluster", which matches the CFT counting with finite particle number correction. While matrix elements to all excited states are shown in (b,d), the small $|I_n|$ sometimes precludes their visualization. We have added a red dashed line wherever necessary to identify the edge counting.}
\label{FigureTwo}
\end{figure}

We start by reviewing the properties of the FQH edge spectrum. In the low-energy limit, the edge modes of FQH states are described by a conformal field theory (CFT), whose nature depends on the topological order in the bulk~\cite{Moore1991Nonabelion, Wen1990ChiralLL, wen1993nonabelian}; for the Laughlin state considered here, it coincides with the chiral Luttinger liquid. This powerful bulk-edge correspondence can be harnessed to identify FQH states from their low-energy spectrum, which reveals the universal counting of the CFT (the number of low-energy edge states for each momentum value).
Extracting the edge spectrum is a non-trivial task, even numerically.
Previous numerical studies have shown that a gapless chiral edge mode, whose counting matches the CFT expectation~\cite{kjall2012edge,luo2013edge}, could be extracted from the energy spectrum in the presence of a smooth confining potential. Interestingly, this property is already present in two-boson systems, with corrections to the CFT counting due to finite particle number~\cite{luo2013edge}\footnote{For 2 particles, the counting for a free chiral boson (1, 1, 2, 3, 5, 7 … ) is truncated down to (1, 1, 2, 2, 3, 3 … )}. This is illustrated in Fig.~\ref{FigureTwo}(a), through the low-energy spectrum of two bosons in the HBH model in a weak harmonic trap, using the eigenvalues of the modified $C_4$ rotation operator~\cite{ozawa} to highlight the chirality of the edge spectrum. 
The situation is different in a box potential, where the absence of low-energy structure prevents the extraction of the edge spectrum; see Fig.~\ref{FigureTwo}(c) and Refs.~\cite{kjall2012edge, repellin2019detecting}. We now show how chiral edge and bulk properties can nonetheless be extracted in these relevant configurations, using a proper spectroscopic probe.

{\it Optical spectroscopy ---} We propose to probe the FQH edge and bulk modes by using two interfering Laguerre-Gaussian (LG) lasers, designed to induce a transition from the prepared FQH ground state to low-energy excitations. This transition involves a transfer of angular momentum $l$, and energy $\hbar \omega$, which are conveniently controlled by the pair of LG beams~\cite{friese} [Fig.~\ref{FigureOne}]. This differs from the detection scheme of Refs.~\cite{dong2018edge,SunPRXQ}, where the dispersion relation is extracted through a Fourier transform of the particle density following a quench. Such a LG-driving scheme was initially proposed in Ref.~\cite{goldman2012edge} to probe the edge spectrum of integer QH states. Beyond the strongly-interacting nature of the FQH state treated in this work, the small system sizes envisaged to realize FQH states in experiments results in a highly discretized spectrum. Our proposed scheme takes these key properties into account, and we discuss their consequences on the resulting absorption spectrum.

The LG modes are solutions of the cylindrical-symmetric wave equation, and take the following form
\begin{equation}
    LG(r,\theta) \propto \left ( \frac{r}{r_0} \right )^{|l|} e^{-\frac{r^2}{2r_0^2}} e^{i\theta l} \equiv f_l(r)e^{i\theta l},
    \label{LaguerreGaussModes}
\end{equation}
where the integer $l$  represents the quantum of orbital angular momentum carried by each photon. As shown in Fig.~\ref{FigureOne}(b), such an optical mode has a spatial vortex structure, with a ring of maximum amplitude that can be adjusted to optimize the edge response. The interference of two such LG beams, with frequencies $\omega_1, \omega_2$ and angular momenta $l_1, l_2$, produces a time-periodic potential acting on the atoms,
\begin{equation}
    \hat O_l(t) = 2\epsilon\sum_j f_l(r_j) \cos{\left (  {\omega t + \theta_j l} \right )}\ \hat b^{\dagger}_j \hat b_j ,
    \label{ProbeOperator}
\end{equation}
where $j$ sums over the lattice sites, $\theta_j$ is the polar angle at $j$, $\omega=\omega_2-\omega_1$ and $l=l_2-l_1$. 

To predict allowed transitions, we calculate the coupling of the FQH ground state $\psi_0$ to excitations $\psi_n$ through the LG drive, i.e. the coupling matrix elements
\begin{equation}
    I_n = \sum_j\mel{\psi_n}{ f_l(r_j)e^{i\theta_j l} \hat b^{\dagger}_j \hat b_j}{\psi_0}.
    \label{TransMatrixElements}
\end{equation}
 We calculate the low-energy eigenstates of $\hat H$ for a system of $N=2$ hard-core bosons on a 10x10 lattice with magnetic flux $\alpha = 1/8$ using exact diagonalization (ED). The corresponding low-energy spectrum and matrix elements $I_n$ are represented in Fig.~\ref{FigureTwo} for a harmonic and a box potential, respectively. In both cases, the matrix elements exhibit a chiral branch at $l < 0$, and two isolated gapped states at $l=1$ and $l=2$. We interpret the gapless $l<0$ branch as the chiral edge mode; as expected, the sign of orbital angular momentum injected by the probe matches the chirality of the edge boson. Conversely, the $l>0$ signal is interpreted as low-energy bulk excitations, which correspond to the MRM~\cite{GMP1986}. While the maximum value of $|l|$ in the edge mode is dictated by the lattice size (here, the 10x10 box), giving rise to a large number of low-energy edge states, the number of MRM states above the Laughlin state is equal to the number of particles ~\cite{repellin2014SMAChern, jolicoeur2017, SuppMat}. The analysis of density profiles (below in the main text, and in~\onlinecite{SuppMat}) corroborates this interpretation of negative and positive $l$ states as edge and bulk states, respectively.
 
 Interestingly, the LG probe identifies the chiral branch and MRM even when no structure can be extracted from the low-energy ED spectrum [Figs.~\ref{FigureTwo}(c) and (d)].
 This behavior originates from the approximate rotation symmetry of the low-energy eigenstates. While the discrete $C_4$ rotation symmetry of our lattice model only guarantees the conservation of the angular momentum $l$ modulo $4$, $l$ is approximately conserved. At small $\alpha$, reduced lattice effects lead to a better conservation of $l$~\cite{SuppMat}. Yet, even for the experimentally relevant regime $\alpha \approx 1/4$, coupling matrix elements that satisfy the conservation of $l \ \mathrm{mod}\ 4$ but not the conservation of $l$ are $40$ times smaller than those that do. Overall, Figs.~\ref{FigureTwo}(b) and (d) show that the addition of a weak harmonic potential increases the velocity of the chiral edge mode, such that its winding becomes visible in the folded energy spectrum [Fig.~\ref{FigureTwo}(a)].

We have chosen the value $r_0 = 2$ for the Gaussian extension of the probe such that $f_l(r)$ remains non-zero both in the bulk and a few magnetic lengths outside the edge of the FQH droplet, at least for $|l| \leq 5$. This is a necessary condition for the corresponding matrix elements to be non-zero, due to the density profiles of the ground state and low-energy excited states, whose spatial extension increases with increasing energy~\cite{SuppMat}. Naturally, the optimal value of $r_0$ depends on the size of the droplet; in larger droplets, it is especially useful to optimize $r_0$ to specific values of $l$~\cite{SuppMat}.

\begin{figure}[t]
\centering
\includegraphics[width=\linewidth]{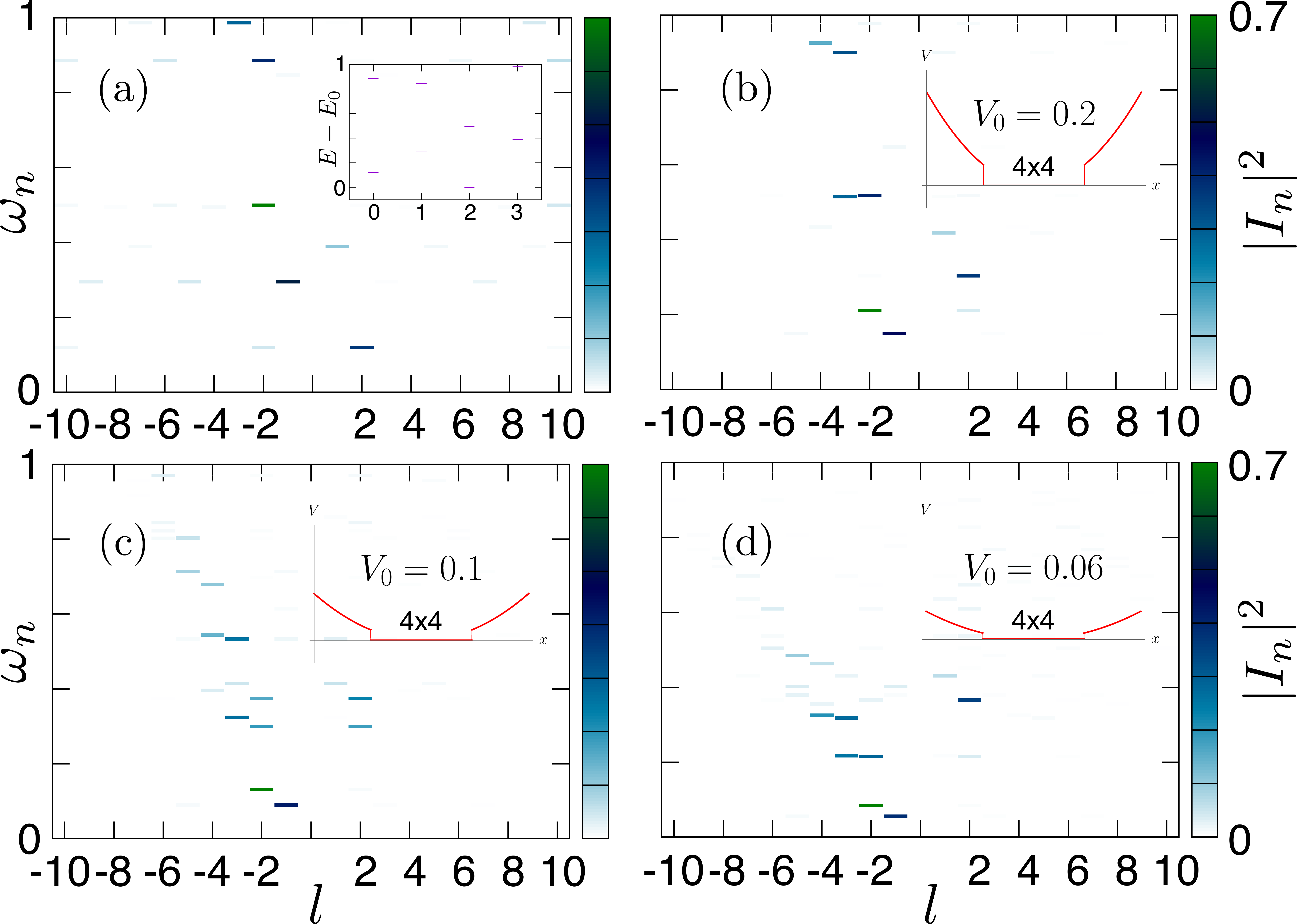}
\caption{Absorption spectra for $N\!=\!2$ interacting bosons ($U=8.1J$) on a 10x10 HBH lattice with flux $\alpha =0.25$, and different trap configurations, keeping $V_0\!=\!0$ in the central 4x4 portion of the lattice. (a) Harvard experiment configuration~\cite{leonard}, with infinite walls around the 4x4 box. The inset shows the corresponding energy spectrum. (b-d) Harmonic potential of respective strength $V_0\!=\!0.2, 0.1, 0.06$. Upon lowering the potential in the outer box, the bulk gap ($l\!=\!2$ signal) increases, while the edge branch ($l<0$) becomes gapless. The weak signal at $l\!=\!2$, at the same energy as the $l\!=\!-2$ edge signal is due to the imperfect conservation of $l$.}
\label{FigureThree}
\end{figure}

{\it Connection to the Harvard experiment} --- We now apply our spectroscopy protocol to the experimental setup of Ref.~\cite{leonard}. There, $N=2$ interacting ($U=8.1J$) bosons are confined to a 4x4 box with hard walls on the HBH lattice, and a FQH ground state was identified within a flux window $0.24 < \alpha < 0.3$. Focusing on $\alpha = 0.25$, we show our probe's matrix elements in Fig.~\ref{FigureThree}(a). The absorption spectrum is consistent with a FQH state whose edge mode is gapped due to the small size of the box. Releasing the walls of the 4x4 box into a 10x10 box confirms this interpretation: the $l<0$ edge branch goes down and becomes gapless, while the $l=2$ bulk gap increases; see Fig.~\ref{FigureThree}, with the trap shape drawn as an inset.
Further numerical investigation~\cite{SuppMat} shows that in a large box, irrespective of the trap shape, the FQH droplet recovers a gapless edge mode, and its bulk gap increases until it reaches its thermodynamic value. Conversely, when the size of the box is reduced, the bulk gap decreases and eventually closes, marking a phase transition. Overall, our calculations show that increasing the size of the quantum-simulation box could permit the observation of a chiral gapless edge mode in ongoing experiments, even in two-atom droplets.

{\it Extracting the edge and bulk modes from in-situ density measurements ---} We now show how to obtain the absorption spectrum studied in the previous paragraphs using observables accessible to cold atom experiments.

\begin{figure}[t]
\centering
\includegraphics[width=\linewidth]{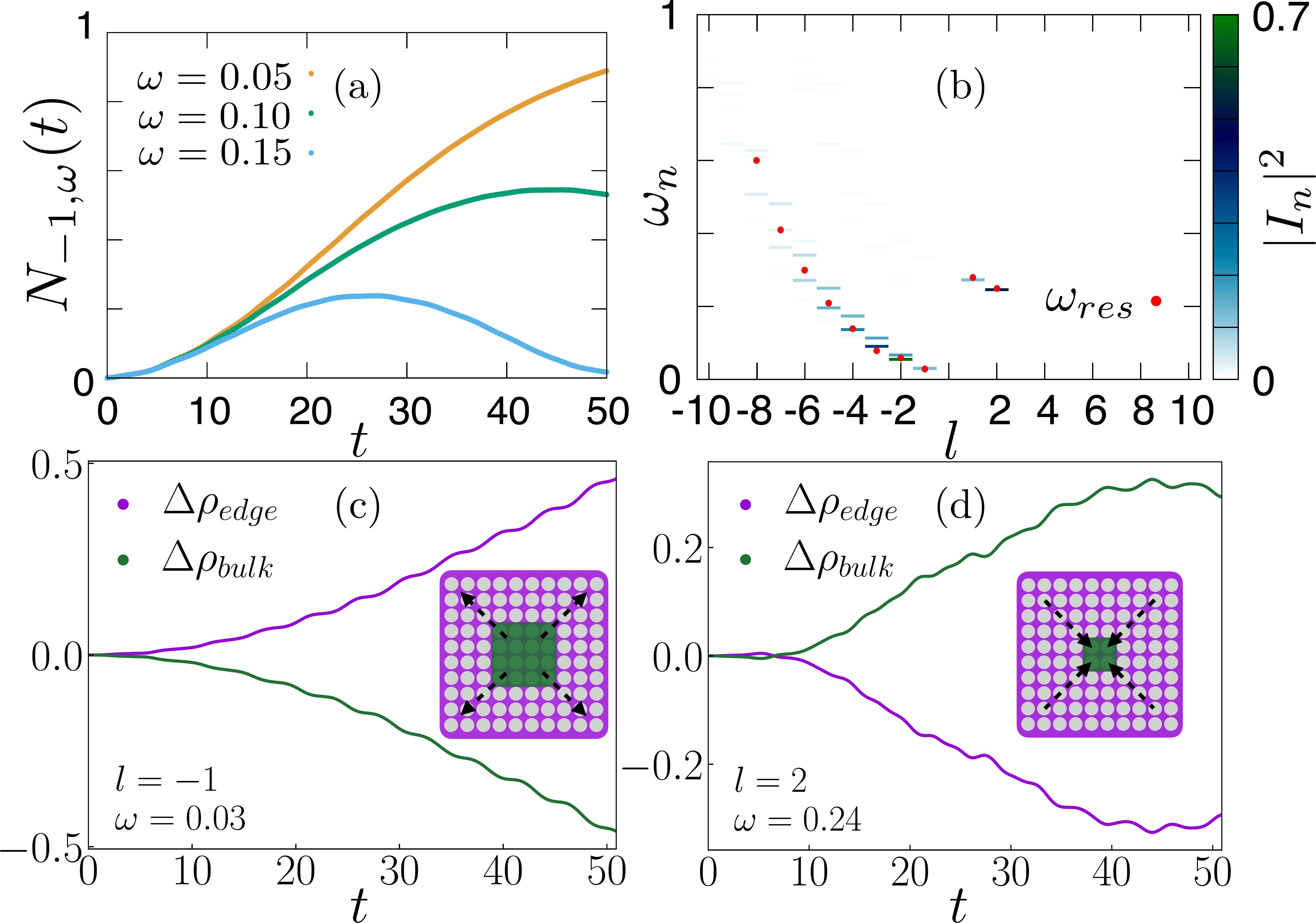}
\caption{Time evolution of 2 hardcore bosons on a 10x10 HBH lattice with flux $\alpha = 1/8$ and a harmonic trap of strength $V_0=0.01$, upon a LG drive of amplitude $\epsilon =0.05$. (a) Excitation fraction $N_{l,\omega}(t) = 1 - |\braket{\psi(t)}{\psi_0}|^2$, for an injected angular momentum $l\!=\!-1$. (b) Chiral branch and magneto-roton mode extracted from the density-transfer protocol (dots). Colored bars indicate the coupling matrix elements. (c)-(d) Density variation at the edge and bulk, for $l\!=\!-1$ and $l\!=\!2$, representative of the typical resonant coupling to edge and bulk states.}
\label{FigureFour}
\end{figure}

Due to its well-defined angular momentum $l$, our LG probe only couples the ground state to a handful of excited states at most, as observed in our analysis of transition matrix elements $I_n$. As a result, the LG drive induces an effective two-level coupling for each value of $l$. Using unitary time-evolution of the FQH ground state subjected to the LG drive, we numerically observe Rabi oscillations, whose amplitude is maximal at the resonance frequency $\omega = \omega_{res}$ [Fig.~\ref{FigureFour}(a)].

We propose to detect the resonant excitations through local density measurements, which are experimentally accessible using a quantum gas microscope~\cite{Tai2017,leonard}. Following the excitation of edge states, we expect the density profile to increasingly populate the external rings outside the bulk. Conversely, we expect transitions to the MRM to result in a density increase in the bulk, due to its nature as a compressibility modulation of the ground state~\cite{SuppMat}
We define the instantaneous edge density as
\begin{equation}
    \Delta\rho_{edge}(t) = \sum_{\substack{j \in edge}} \rho_j(t) - \rho_j(0) , \label{RhoEdgeObservable}
\end{equation}
where $\rho_j$ is the density on site $j$, and we have defined the complementary bulk and edge regions [green and purple regions in the inset of Fig. \ref{FigureFour}(b)] from the ground state density profile~\cite{SuppMat}. The time-dependent behavior of $\Delta\rho_{edge}(t)$, as obtained using numerical time-evolution in Fig.~\ref{FigureFour}(b), confirms the migration of the particle density from bulk to edge (respectively edge to bulk) for a negative (respectively positive) injected angular momentum. Figure~\ref{FigureFour}(c) shows the values of $\Delta\rho_{edge}$ at different probing frequencies $\omega$, taken after an observation time $t^*=50\hbar/J$: we can clearly distinguish the resonance around $\omega=0.03$ for $l=-1$. Iterating this procedure for different values of the injected angular momentum $l$, we retrieve the chiral edge spectrum and MRM in Fig. \ref{FigureFour}(d). We point out that a realistic observation time $t^* \sim 10 - 100 \hbar/J$ leads to a clear signal $\Delta\rho_{edge}(t^*)\sim 0.1-1$, which can be detected using a quantum gas microscope~\cite{leonard}. We also verified the ability of our protocol to detect edge signals in the experimental configuration of Ref.~\cite{leonard}~\cite{SuppMat}. Lastly, performing a full transfer to a target edge state would allow to image its hallmark vortex structure in-situ~\cite{SuppMat}.

{\it Discussion ---} Our work indicates that the chiral edge branch and MRM of atomic FQH droplets can be probed by measuring the local density following a Laguerre-Gaussian drive. Our method is well suited to probe the few-atom droplets addressed by ongoing experiments. Increasing the lattice size beyond the small boxes realized so far in experiments~\cite{leonard} would permit the extraction of a gapless edge mode even in two-particle systems. We have verified the validity of our results beyond this limit, for $3$ or $4$ particles, where we can still address very dilute systems~\cite{SuppMat}.
Beyond experimental purposes, the calculation of LG matrix elements is a convenient tool to analyze bulk and edge properties of FQH states, especially in lattice systems without continuous rotation symmetry. It distinguishes bulk and edge states, allowing e.g. the tracking of the bulk gap and of the velocity of the edge branch upon changing the confining conditions.

Finally, our method is promising in view of identifying the fractional statistics of anyonic excitations (see also  Ref.~\onlinecite{cooper2015statistics}). Indeed,
the number of edge states detected by our LG probe for each value of the angular momentum $l$ matches the CFT counting for a free chiral boson (with corrections due to the small particle number~\cite{luo2013edge}), which describes the edge of a Laughlin droplet. We have verified that a tighter confinement of the FQH droplet leads to larger energy splittings, permitting the resolution of resonances occurring at the same $l$ without requiring longer observation times~\cite{SuppMat}. Overall, our scheme allows us to extract the edge counting at least up to $l\!=\!-3$, which is enough to distinguish most FQH states~\footnote{The counting is respectively (1, 1, 2, 3) and (1, 1, 3, 5) for the Laughlin and Moore-Read states.} The MRM constitutes an additional topological signature, absent from integer quantum Hall phases, and whose maximum angular momentum distinguishes between different FQH phases~\footnote{$L_{max} = N$ and $L_{max}=N/2$ are respectively expected for Laughlin and Moore-Read}.

{\it Acknowledgments ---} We thank Brice Bakkali-Hassani, Jean Dalibard, Andr\'e Eckardt, Markus Greiner, Fabian Grusdt, Adolfo G. Grushin, Markus Holzmann, Joyce Kwan, Julian L\'eonard, Yanfei Li, Felix Palm, Perrin Segura and Botao Wang for insightful discussions. C.R. acknowledges support from ANR through grant ANR-22-CE30-0022-01. Work in Brussels is supported by the FRS-FNRS (Belgium), the ERC Starting Grants TopoCold and LATIS, and the EOS project CHEQS.

\bibliography{edge_FCI}

\newpage
\onecolumngrid

\newpage
\begin{center}
\textbf{\large Supplemental material for: \\ Spectroscopy of edge and bulk collective modes in fractional Chern insulators}
\end{center}

\twocolumngrid

\section{Influence of lattice effects on angular momentum selection rules} \label{AppSelectionRules}
\begin{figure}[t]
\begin{minipage}[b]{\linewidth}
\centering
\includegraphics[width=\textwidth]{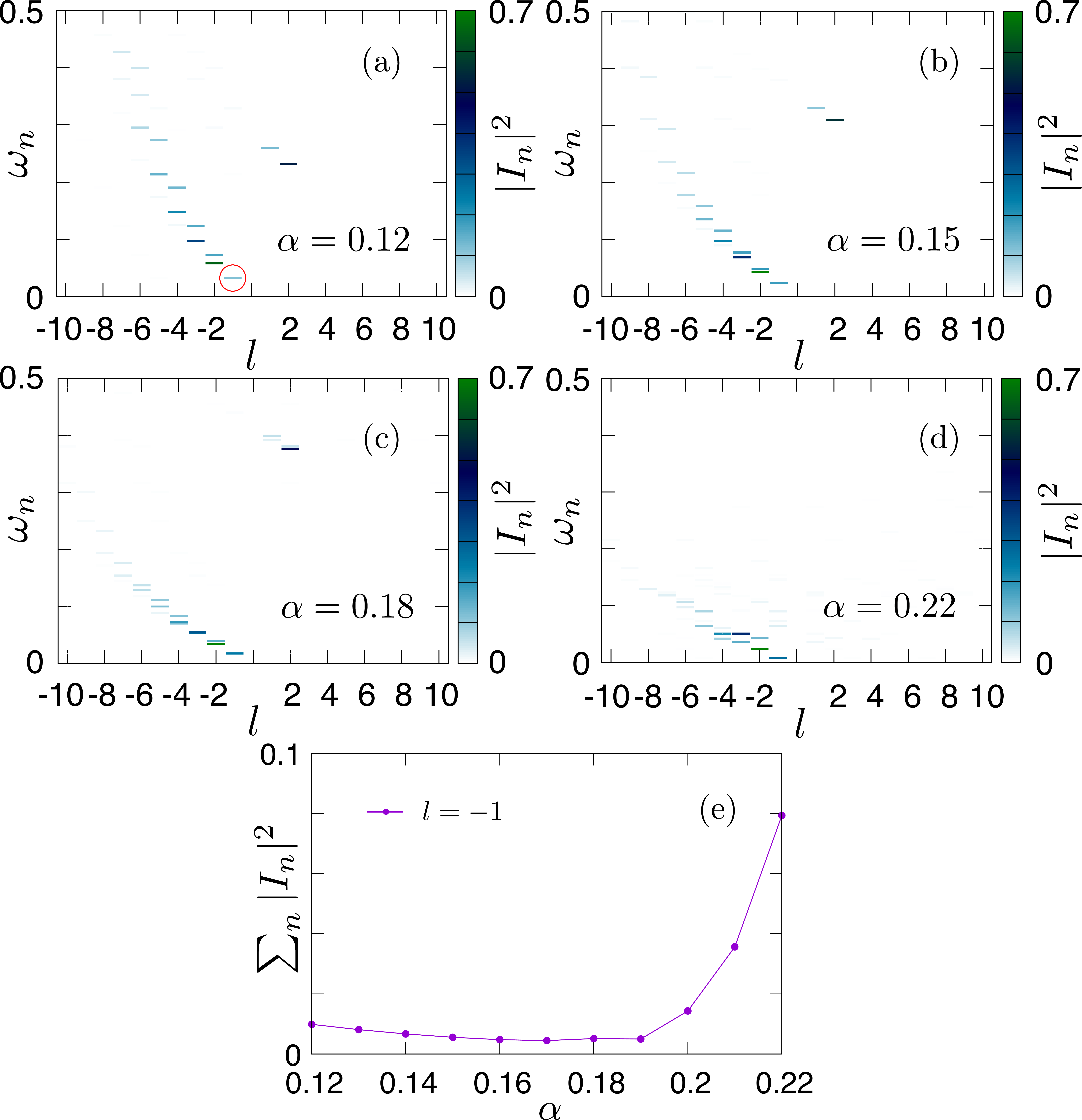}
\end{minipage}
\caption{(a-d) Coupling matrix elements of the LG probe for a system of $N=2$ bosons on a harmonically trapped ($V_0=0.01$) 10x10 lattice with different flux densities $\alpha$. The LG beam has a radius $r_0=2$. (e) Sum of matrix elements in the sectors $l=-1$ as a function of the flux density, excluding the red-circled, lowest energy state.}
\label{FigureA}
\end{figure}

In the continuum, the conservation of continuous rotation symmetry results in a selection rule for the coupling matrix elements $I_n$ defined in Eq.~(4) of the main text, which would allow the detection of a chiral edge branch with unbounded angular momentum in the absorption spectrum. In a square lattice with discrete rotation symmetry, in principle, we expect non-zero matrix elements $I_n = \mel{\psi_{L'}}{\hat O_l}{\psi_L}$ whenever $L' = L + l (mod) 4$, where $L$ and $L' \in \mathbb{Z}_4$ are the rotation eigenvalues of two Hamiltonian eigenstates. Yet, as we have observed in the main text, many matrix elements satisfying this discrete rotation selection rule are negligible. This reveals the approximate continuous rotation symmetry, which becomes exact in the continuum limit $\alpha \ll 1$ of the Hofstadter Hamiltonian. To further illustrate this phenomenon, we show the evolution of matrix elements upon increasing the flux from $\alpha =0.12$ to $\alpha = 0.22$ in Fig. \ref{FigureA}. We note that the ground state remains a FQH state across this magnetic flux range, as denoted by the visible chiral edge branch at $l<0$ and bulk gap at $l=2$.
Figure \ref{FigureA}(e) shows the sum over excited states of the $l=-1$ matrix elements; it highlights the increase in amplitude of the spurious signal associated with $l=-1 + 4p$, where $p$ is a non-zero integer (the biggest contribution comes from $l=-5$, as is visible in Fig. \ref{FigureA}(d)).

\section{Time-dependent density measurements in the Harvard-experiment configuration} \label{AppExperimentalSetup}
\begin{figure}[t]
\begin{minipage}[b]{\linewidth}
\centering
\includegraphics[width=\textwidth]{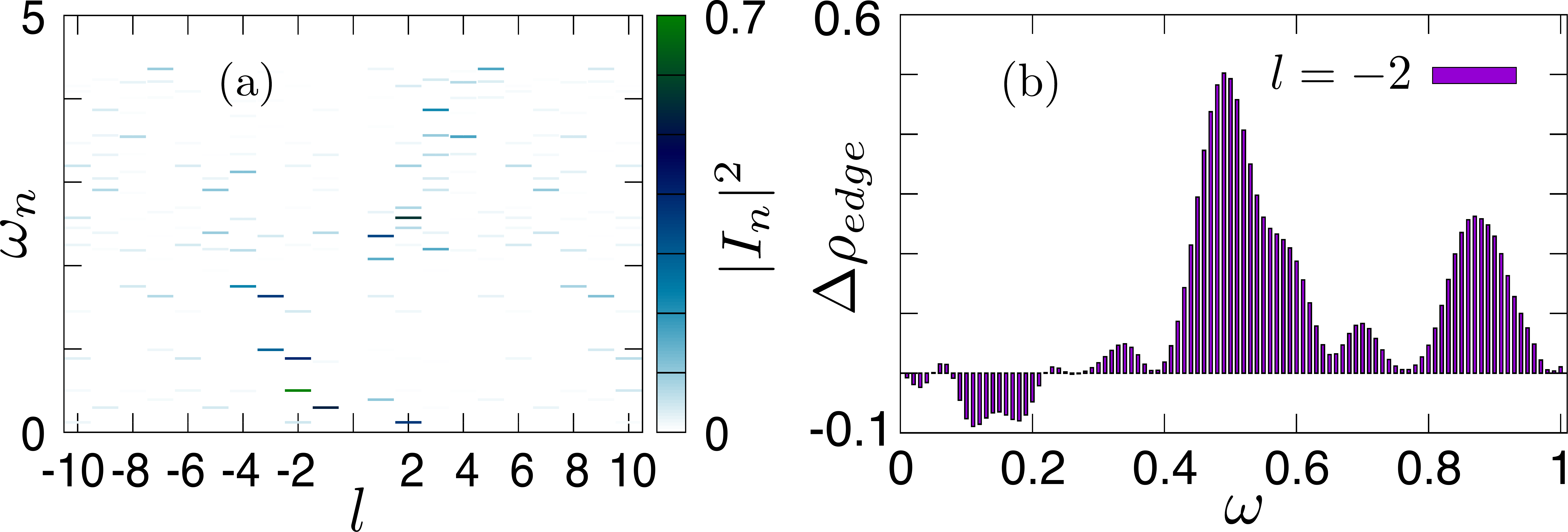}
\end{minipage}
\caption{(a) Absorption spectrum for $N=2$ bosons on a 4x4 lattice with hard walls ($V_0=0$), for a flux density $\alpha=0.25$ and on-site interaction $U=8.1J$. The laser beam has a radius $r_0=2$. (b) Bulk to edge density transfer, as retrieved through the time-dependent protocol explained in the main text, probing the $l=-2$ sector of the absorption spectrum in (a), showing the absence of bulk to edge transfer below $\omega \simeq 0.4$}
\label{FigureB}
\end{figure}

In the main text we presented the absorption spectra related to the experimental setup of Ref.~\cite{leonard}. We now apply our time-dependent protocol to test its robustness in real scenarios. We show again the absorption spectrum in the experimental configuration in Fig. \ref{FigureB}(a) but this time on an extended energy scale.
We note the presence of spurious states, associated with the imperfect conservation of angular momentum $l$ in this small lattice system. For example, the low-energy state previously identified as the bulk gap is present with its largest matrix element at $l=2$, but it is also visible (with a much smaller amplitude) at $l=-2$ and $l=\pm10$. We might then worry that this spurious bulk signal at $l=-2$ might obfuscate the extraction of the edge spectrum. Fortunately, our time-dependent density measurement protocol easily distinguishes edge and bulk signal, as is shown in Fig.~\ref{FigureB}(b), since the bulk-to-edge density transfer only occurs when the LG probe couples the ground state to an edge state.

\section{Exploring the dilute to dense transition} \label{AppChiralBranch}
In the main text we showed how a gapless edge mode can be detected in the absorption spectrum when we have a sufficiently dilute system (i.e. when the simulation box is larger than the FQH droplet). On the other hand we have seen how the chiral branch becomes gapped if the box size nearly coincides with the size of the FQH droplet, as in the experimental configuration~\cite{leonard} involving $N=2$ bosons in a 4x4 box. 

\begin{figure}[t]
\begin{minipage}[b]{\linewidth}
\centering
\includegraphics[width=\textwidth]{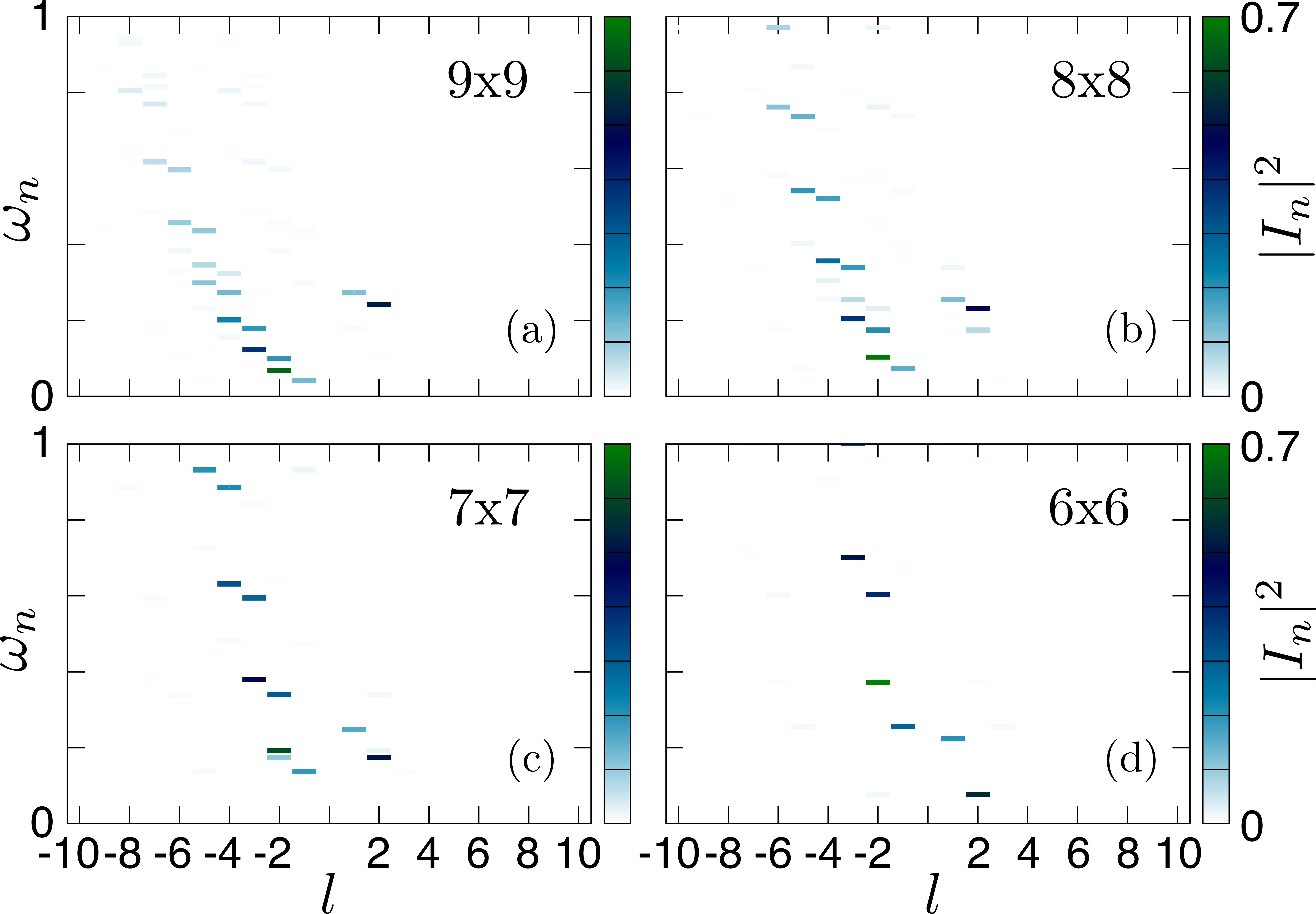}
\end{minipage}
\caption{(a)-(d) Absorption spectra for different lattice size, in the case of $N=2$ bosons and a flux density $\alpha=0.125$, in a harmonic trap ($V_0=0.01$) and a LG radius $r_0=2$. We observe a transition to a trivial phase for a 5x5 lattice (not shown).}
\label{FigureC}
\end{figure}

In Fig.~\ref{FigureC}, we show additional data illustrating this phenomenon, for hard wall boxes of different sizes, a droplet of $N=2$ bosons and a confinement potential of strength $V_0=0.01$, with a flux density $\alpha=0.125$. We observe the chiral branch becoming steeper as the number of sites decreases, and the bulk gap (as tracked through the $l=2$ signal) decreasing accordingly. The closure of the bulk gap occurs between the 6x6 and the 5x5 boxes. We have verified that the ground state in the 5x5 box is a trivial state, which is adiabatically connected to the infinite trap limit (particles confined to the central site of the lattice).

Note that Fig.~\ref{FigureA} illustrates the same phenomenon: there, the size of the FQH droplet changes, while the box size does not. As $\alpha$ decreases, the radius of the FQH droplet increases (so that the density in the bulk satisfies Streda formula~\cite{repellin2020drift}), which results in a steeper edge mode and smaller bulk gap.

\section{Local density patterns} \label{AppLocalDensity}
\begin{figure}[t]
\begin{minipage}[b]{\linewidth}
\centering
\includegraphics[width=\textwidth]{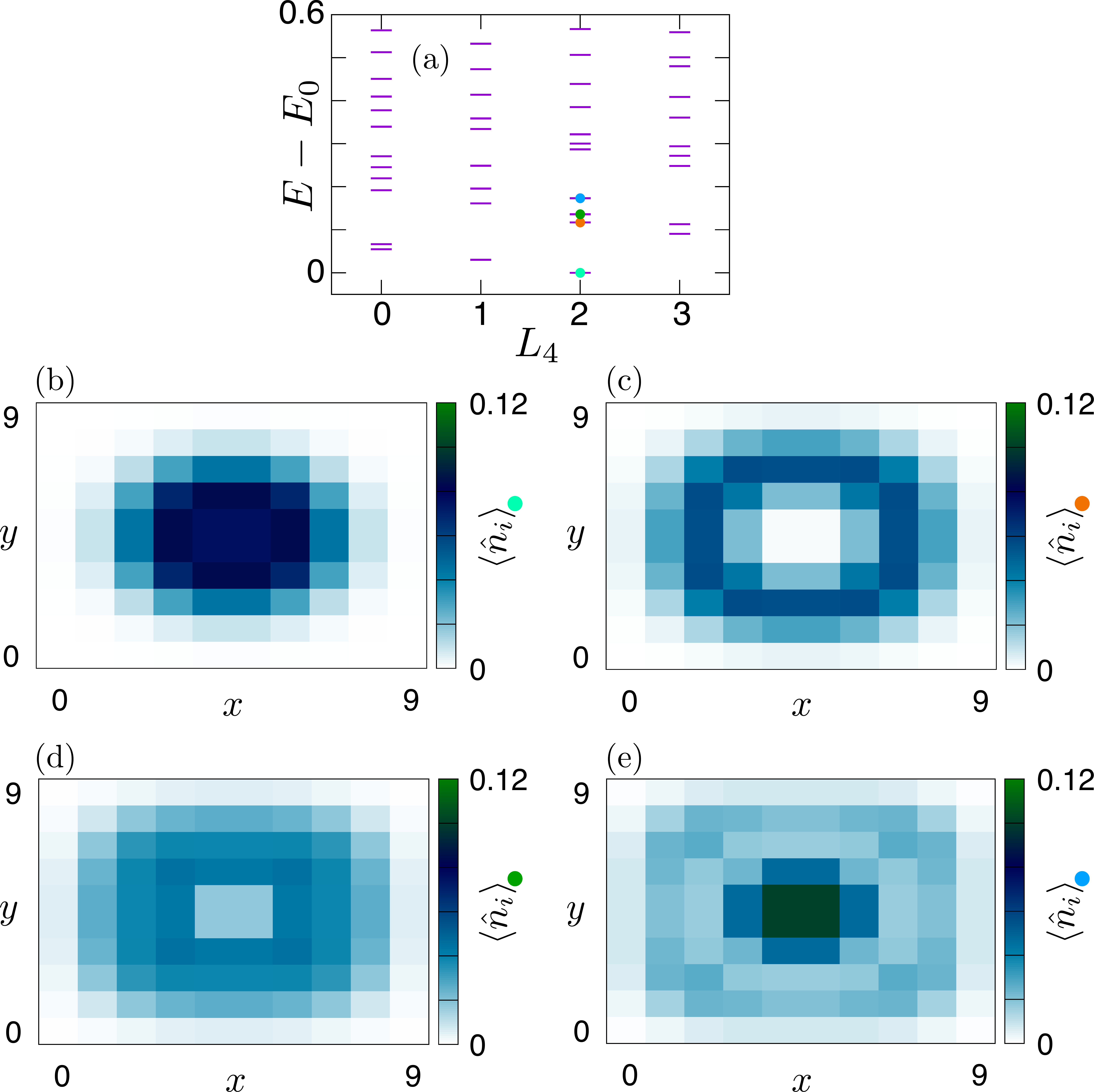}
\end{minipage}
\caption{(a) Energy spectrum for $N=2$ bosons in a 10x10 lattice, for a flux density $\alpha=0.125$ and a harmonic trap of strength $V_0=0.01$. The four coloured dots represent the state considered for the local density calculations in the next panels. (b) Local density profile of the ground state. (c)-(e) Local density profiles of three different excited states.}
\label{FigureD}
\end{figure}

Here, we show the local density of the ground state and edge excitations in our model. We consider $N=2$ bosons on a 10x10 square lattice with flux density $\alpha=0.125$, confined in a harmonic trap of strength $V_0=0.01$. The ED spectrum in this configuration was shown in the main text, and is reproduced in Fig. \ref{FigureD}(a). The eigenstates considered for the calculation of the density are marked with coloured dots. The ground state density profile in Fig. \ref{FigureD}(b) shows a FQH droplet mostly confined to the center square, whereas the density of the excited states in Fig. \ref{FigureD}(c)-(e) is more spread out. This behaviour explains the success of our detection protocol based on density measurement.

\section{Absorption spectra for $N=3,4$ particles: chiral branch and magneto-roton mode}\label{AppMR}
\begin{figure}[t]
\centering
\includegraphics[width=\linewidth]{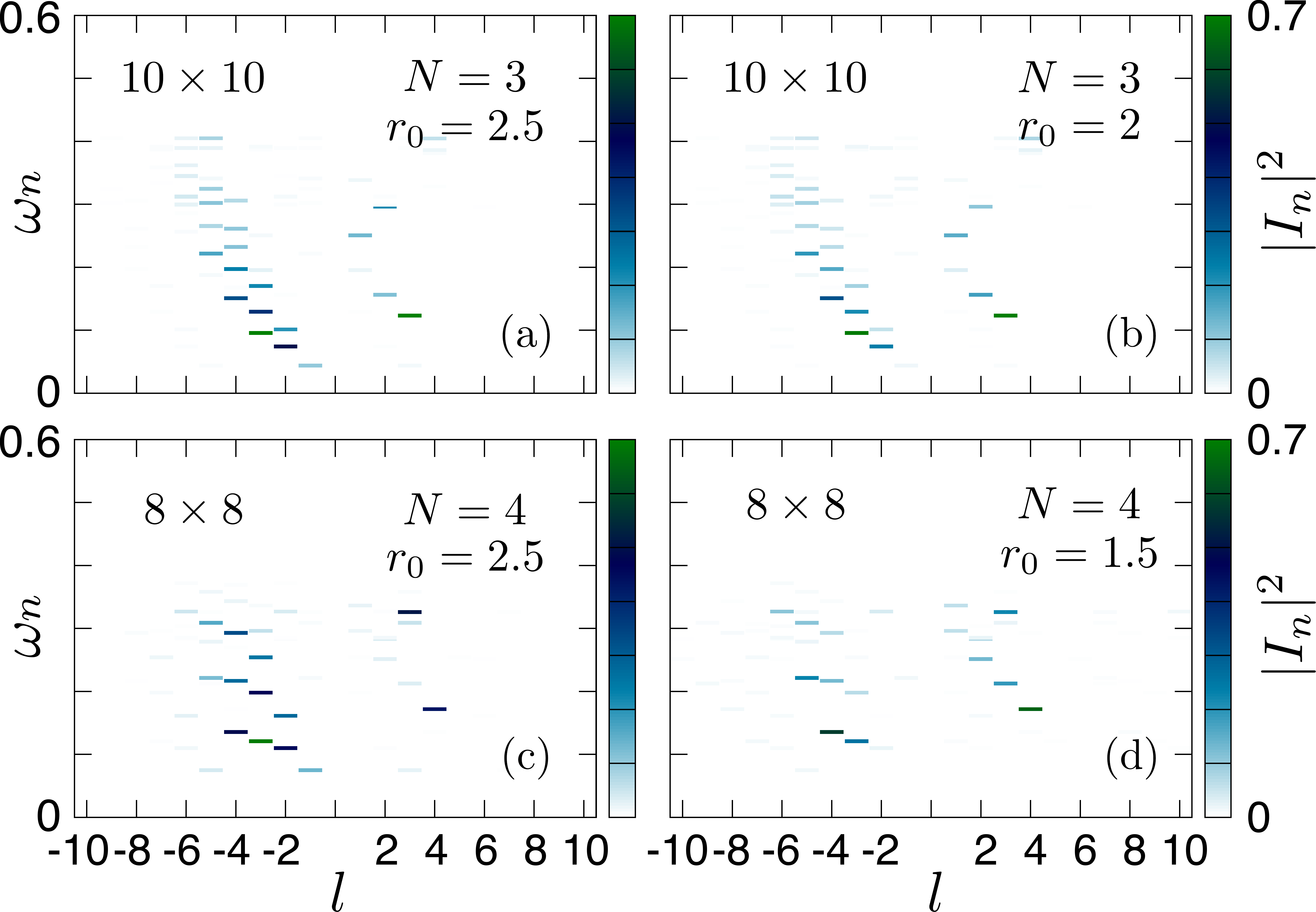}
\caption{Absorption spectra at different LG beam radii $r_0$ for $N=3$ bosons in a $10 \times 10$ lattice with a flux density $\alpha=1/8$ (a,b) and $N=4$ bosons in a $8 \times 8$ lattice with a flux density $\alpha=1/5$ (c,d) trapped in a harmonic potential of strength $V_0=0.01$. We emphasize the chiral branch in the $l<0$ region when $r_0=2.5$ for $N=3,4$ particles (a,c), whereas we show the emergent magneto-roton mode in the $l>0$ region when the LG beam is acting deep in the bulk for low $l$ values, namely when $r_0=2$ for $N=3$ (b) and $r_0=1.5$ for $N=4$ (d).}
\label{FigureMR}
\end{figure}

We now verify the validity of our protocol in systems with $N=3$ and $N=4$ particles, which are amenable to ED calculations in a lattice of respective size 10x10 and 8x8. A gapless edge branch requires the walls of the simulation box to lie sufficiently far outside of the FQH droplet, which prevents us from addressing systems with more than $4$ particles with ED. For $N=3$, we have used a flux density $\alpha=1/8$, while for $N=4$, we have used a larger value $\alpha=1/5$ to decrease the radius of the FQH droplet in this smaller box. In Fig.~\ref{FigureMR}, we can identify the chiral edge branch in the $l<0$ region, in both cases.
Let us now focus on the bulk excitations in the $l>0$ region. We find a low-energy mode with a cut-off at $l_{max} = N$, in agreement with our interpretation of this branch as the magneto-roton mode~\cite{repellin2014SMAChern, jolicoeur2017}.

The size of the FQH droplet scales with the number of particles $N$, so it is reasonable to expect an optimized chiral response by increasing the LG beam radius $r_0$ compared to the $N=2$ case of the main text. In Fig. \ref{FigureMR}(a,c) the edge states are probed with the same $r_0=2.5$. Note that in order to emphasize the bulk response, it is useful to adjust the value of $r_0$ to the value of injected angular momentum $l$; specifically, we tuned $r_0=2.0$ in order to have low-angular momentum injections acting mostly in the bulk region. This is especially important in systems where the FQH droplet is larger (larger number of particles in particular).

\section{Probing the magneto-roton mode through density measurements}\label{AppBulkProbe}
As emphasized in the main text, the magneto-roton mode (MRM) is a unique feature of FQH states, whose maximum angular momentum $L_{\mathrm{max}}$ depends on the nature of the FQH state. Figure~\ref{FigureMR} and Fig.~4 from the main text numerically show that $L_{\mathrm{max}} = N$ for numbers of particle $N=2, 3, 4$, as expected for a Laughlin 1/2 ground state.

\begin{figure}[t]
\begin{minipage}[b]{\linewidth}
\centering
\includegraphics[width=\textwidth]{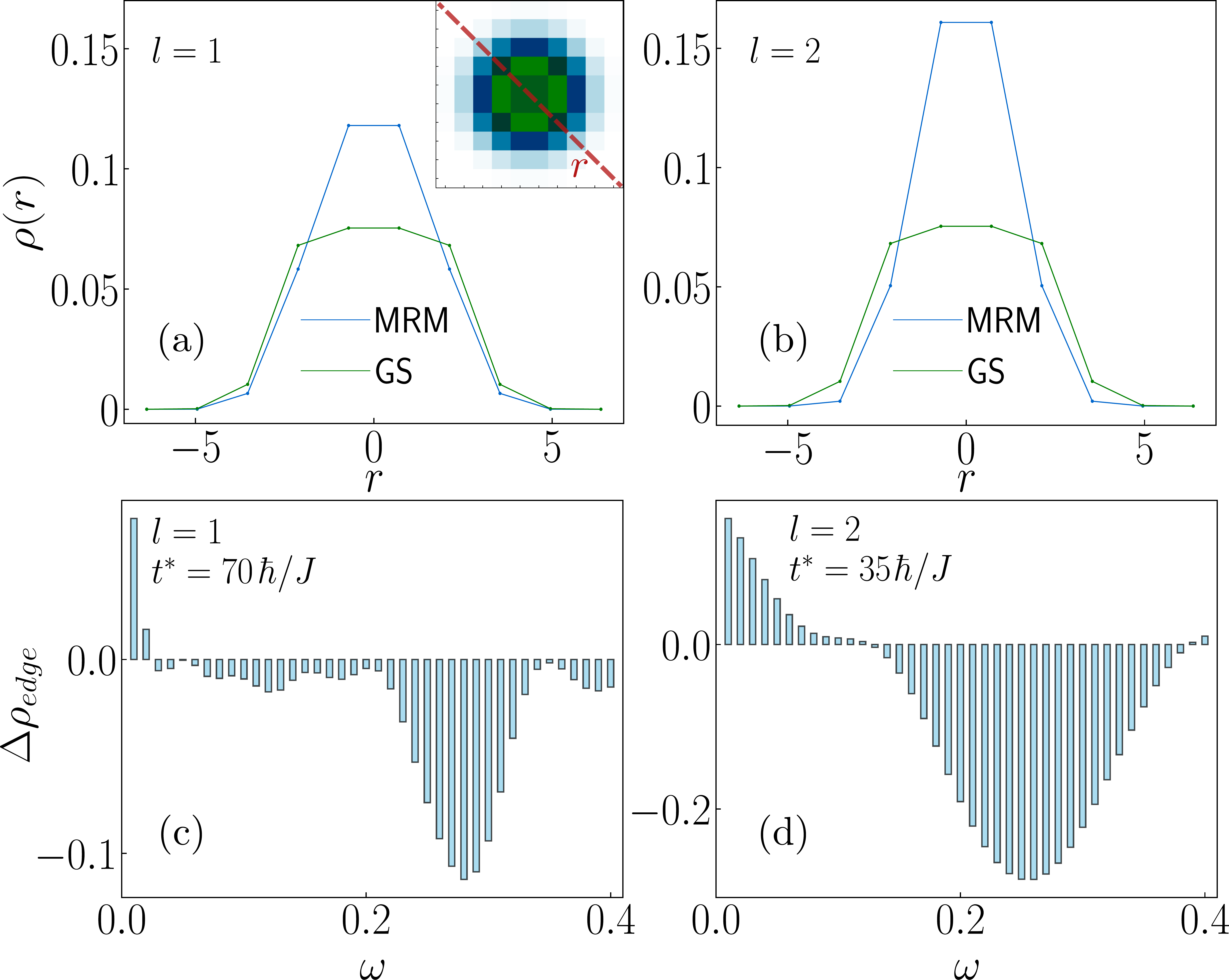}
\end{minipage}
\caption{Probing the magneto-roton mode by monitoring the edge to bulk density transfer for $N=2$ bosons on a $10 \times 10$ lattice, $V_0=0.01$ and $\alpha=1/8$. (a)-(b) Particle density along a diagonal cut of the lattice (red dashed line shown in the inset) for the two magneto-roton mode (MRM) eigenstates (with angular momentum $l=1$ and $l=2$ relative to the ground state) and the FQH ground state (GS). (c)-(d) Bulk to edge density increment at observation time $t^*$. The negative peaks indicate resonant transitions to the magneto-roton states.}
\label{FigureE}
\end{figure}

We now show additional numerical evidence illustrating how our protocol can be applied to extract bulk excitations at $l>0$, and thus extract the MRM of FQH states. 
In Fig.~\ref{FigureE}(a,b) we show the density profile of the FQH ground state (GS) and first bulk excited states (MRM) ($l=1, 2$ in our 2-particle system). The density in the center of the droplet is significantly larger in the MRM than in the ground state.
By driving the ground state with the LG beams at $l=1,2$, we generate a Rabi oscillation to MRM states (similarly to the Rabi oscillations to edge states illustrated in Fig.~4a). This results in a transfer of density from the outer part of the droplet  towards its center. Specifically, in our $2$ particle system on a $10\times10$ lattice, we find that monitoring the density increment in the $2 \times 2$ central region of the lattice unambiguously detects the position of the resonance frequency $\omega_{res}$ as shown in Fig.~\ref{FigureE}(c,d).

\section{Resolving the edge counting}\label{AppProbeCounting}
\begin{figure}[t]
\begin{minipage}[b]{\linewidth}
\centering
\includegraphics[width=\textwidth]{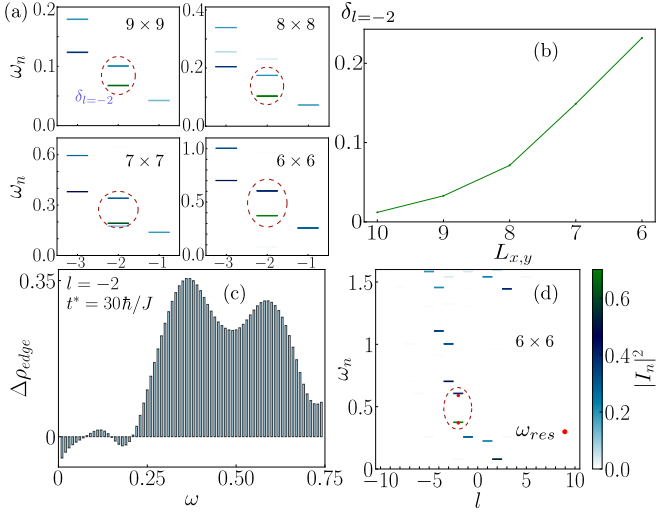}
\end{minipage}
\caption{Resolving resonances occurring at the same $l$ in the edge spectrum. (a) Zoomed absorption spectra emphasizing the edge states at $l=-2$ (red dashed circle) and their energy splitting $\delta_l$ for different lattice sizes $L_x \mathrm{x} L_y$ ($N=2$ bosons, $\alpha=0.125$, $V_0=0.01$). (b) Energy splitting between the two states at $l=-2$, as a function of lattice size $L_{x,y}$. (c) Bulk to edge density increment at observation time $t^*$, showing two well separated resonance peaks. (d) Absorption spectrum with the resonance points (red dots) obtained through our protocol.}
\label{FigureF}
\end{figure}

As emphasized in the main text, the edge state counting is a universal hallmark of topological order, which distinguishes different FQH phases. To access this counting, it is essential to resolve the multiple resonances occurring at the same value of $l$. Here, we show that the energy splitting $\delta_l$ may be enhanced through a judicious choice of confining trap. Specifically, reducing the size of the lattice box $L_x \times L_y$ results in an increased $\delta_l$ [Fig.~\ref{FigureF} a, b]. Consequently, a small (experimentally realistic) observation time is enough to resolve the multiple resonances [Fig.~\ref{FigureF} c,d]. Note that the energy splitting increases with $|l|$, such that being able to resolve the two states at $l=-2$ implies the possibility to resolve states at $l \leq -3$. Naturally, if the lattice box size is too small, the bulk gap closes, as we have discussed in the main text and in the section 'Exploring the dilute to dense transition'. The parameters chosen in Fig.~\ref{FigureF} c,d represent a compromise between optimizing the edge state splitting and keeping the bulk gap open.

\end{document}